\documentclass[twocolumn,superscriptaddress,showpacs,aps,amsmath,amssymb]{revtex4}
\usepackage{bm}
\usepackage{graphicx}
\newcommand{\nl}{\nonumber \\}

\newcommand{\ep}{\epsilon}
\newcommand{\w}{\omega}

\newcommand{\be}{\begin{equation}}
\newcommand{\ee}{\end{equation}}
\newcommand{\bea}{\begin{eqnarray}}
\newcommand{\eea}{\end{eqnarray}}
\newcommand{\bsube}{\begin{subequations}}
\newcommand{\esube}{\end{subequations}}

\newcommand{\Fig}[1]{Fig.\,\ref{#1}}

\newcommand{\comments}[1]{}

\begin{document}

\title{Long-range Entanglement of Kondo Clouds in Open Triple Quantum Dots}

\author{YongXi~Cheng}
\affiliation{Beijing Computational Science Research Center, Beijing 100084, China}
\affiliation{Department of Science, Taiyuan Institute of Technology, Taiyuan
030008, China }

\author{YuanDong~Wang}
\affiliation{Department of Physics, Renmin University of China, Beijing 100872, China}

\author{JianHua~Wei}\email{wjh@ruc.edu.cn}
\affiliation{Department of Physics, Renmin University of China, Beijing 100872, China}

\author{Hong-Gang~Luo}
\affiliation{School of Physical Science and Technology $\&$ Key Laboratory for Magnetism and Magnetic Materials of the MoE, Lanzhou University, Lanzhou 730000, China}
\affiliation{Beijing Computational Science Research Center, Beijing 100084, China}

\author{HaiQing~Lin}
\affiliation{Beijing Computational Science Research Center, Beijing 100084, China}

\date{\today}

\begin{abstract}
We study entanglement of Kondo clouds in an open triple quantum dots (OTQDs) system based on the dissipaton equation of motion (DEOM) theory.
A comprehensive picture of the long-range entanglement of Kondo clouds is sketched by the spectral functions, spin-spin correlation and dot occupancies of OTQDs.
We find that for the configuration $(N_{1},N_{2},N_{3})=(1,0,1)$, a conduction electrons peak occurs in the spectral function of intermediate QD in Kondo regime.
This peak resulting from the overlapping of the two Kondo clouds forming from between the two peripheral QDs and leads, enhances with decreasing temperature and increasing  dot-lead coupling.
Both the spin-spin correlations between the two adjacent QDs and the two peripheral QDs owns negative values.
It also confirms the physical picuture of the overlapping between left and right Kondo clouds via the intermediate QD.
Moreover, the transition of the effective electron occupation and the spectral function of intermediate QD in Kondo regime also indicates the entanglement of Kondo clouds enhancing with decreasing temperature and increasing dot-lead coupling.
This investigation will be beneficial to detect the Kondo clouds and to further explore Kondo physics in related experiment setups.
\end{abstract}

\pacs{73.63.Kv, 85.35.Be, 72.15.Qm}  %, 71.10.Fd}

\maketitle
\section{INTRODUCTION}
\label{Intro}

Triple quantum dots system (TQDs) presents a rich new physical phenomena  and indicates a potential device applications due to the extended freedom of couplings and geometric arrangement.
As one of the most simplest devices, TQDs provides an ideal platform for investigating the coded qubit \cite{2012nn54,2013nn654}, frustration \cite{2013prl046803} and quantum teleportation \cite{2012rpp114501}.
Recently, more interesting physics such as Bipolar spin-blockade effects\cite{2013nn261}, complex coherent interactions \cite{2016nature188,2012nn54} and long-range transport \cite{2013nn432,2014prl176803} has been detected on TQDs device.
Moreover, the investigation of the TQDs is the first step to study the multiple-``impurity" problems involving quantum phase transitions \cite{2006prb045312} and Kondo effect \cite{2006prb153307,2013prl047201} in such a system.

As a typical example of quantum many body effect, the Kondo effect results from the screening of a localised spin by the surrounding conduction electrons, where a Kondo spin singlet state is formed in the system \cite{1964ptp37,1993cambrige}.
This effect in QDs system has been demonstrated as a pronounced zero-bias conductance peak at the temperatures below a characteristic Kondo temperature \cite{1998Science540,2000nature342}.
Recently, a vast amount of theoretical works have been carried out to study the Kondo effect in TQDs. The fascinating physics such as two-channel Kondo effect \cite{2010prb075126,2011prb035119}, ferromagnetic and antiferromagnetic Kondo physics phases \cite{2009prb085124,2013prl047201} and non-Fermi-liquid(NFL) behavior \cite{2007prl047203} are explored in TQDs system.

As one of the present main foci, the Kondo clouds comprising the impurity spin and a cloud of spin-correlated electrons surrounding, attract great interest both theoretically and experimentally.
Researchers have been looking intensively for ways of observing the Kondo clouds.
The long-range Kondo signature for extensional Kondo clouds in single magnetic atoms of Fe and Co buried under a Cu(100) surface has been observed experimentally \cite{2011nature203}.
The spatial distribution of the Kondo screening clouds in the Kondo impurity \cite{2008prb104401}, 3-D metal case \cite{2007axiv} and single QD embedded in mesoscopic closed ring \cite{2001prl2854} is characterized.
It is found that the hole coherence factor of the Kondo-doublet interaction and persistent current through the system depends strongly on the the Kondo screening clouds.
The characteristic Kondo length scale of Kondo screening clouds in the Kondo model has also been evaluated theoretically \cite{2010axiv2209}. Due to related the Kondo temperature, the Kondo length scale can be introduced as $\xi_k=\hbar v_{F}/(k_{B}T_{K})$ with $v_{F}$ is the Fermi velocity.
Moreover, the physical picture of the dynamic Kondo clouds \cite{2004jmmm272} and polarization Kondo clouds in Kondo systems \cite{2010epjb95} are investigated.
A Friedel oscillations correlated with the Kondo length is presented.
Recently, Lee \emph{et al}. also have studied the macroscopic quantum entanglement of a Kondo clouds via detecting the entanglement entropy \cite{2015prl057103}.
More importantly, the long-range entanglement of the Kondo screening clouds and the local moment for Kondo impurity model are characterized by studying the entanglement entropy \cite{2018prl147602}.
However, the structure of Kondo screening clouds and the role of Kondo clouds scale in the TQDs system have not a clear theoretical description.
The macroscopic entanglement would be a direct tool for characterizing the properties of the Kondo clouds.
Here, we will study the long-range entanglement of Kondo clouds based on TQDs model.
Indeed, varied configurations of the TQDs arrangements \cite{2006prl036807,2007prb075306,2010prb075304,2008prb193306,2008apl013126,2009apl092103} fabricated experimentally offer new opportunities to clarify the structure of the Kondo screening clouds as well as its role in the interaction between QDs.

\begin{figure}
\includegraphics[width=0.95\columnwidth]{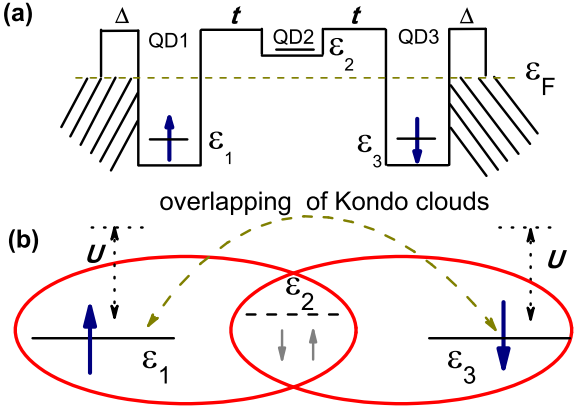}
\caption{ (a) Schematic representation of the open triple quantum dots system (OTQDs).
The TQDs is embedded between two leads (Source and Drain) via the dot-lead coupling strength $\Delta$.
In the present work, QD1 and QD3 are symmetric and both in the localized momentum regime with $N_1=N_3=1$.
While QD2 is nearly unoccupied as $\varepsilon_2 > \varepsilon_F$.
(b) The overlapping of Kondo clouds leads an effective electron occupation on QD2 and a long-rage correlation between QD1 and QD3.}
\label{fig0}
\end{figure}

In this paper, we study the long-range Kondo signature and extensional Kondo clouds via the open triple quantum dots system (OTQDs) based on the dissipaton equation of motion (DEOM) theory\cite{2014jcp054105}, a quasi-particle generalization of the hierarchical equation of motion
(HEOM) formalism \cite{Jin08234703,2012prl266403,2015njp033009}.
Unlike the classical studying for the properties of Kondo clouds structures itself depending upon the distance to the magnetic impurity \cite{2011prb115120,2015prb085127,2010prb045111},
we focus on the transition of the spectral functions, dot occupancies of each QD and spin-spin correlation between related QDs in OTQDs, which can effectively describe the properties of Kondo clouds.
A new schema for studying kondo cloud is proposed, which is easy to implement in the experimental device.
The geometry is depicted as Fig.~\ref{fig0} (a), two peripheral QDs (QD1 and QD3) are directly coupled to the leads via the dot-lead coupling strength $\Delta$, and the intermediate QD (QD2) symmetrically couples to QD1 and QD3 via the interdot coupling strength $t_{12}=t_{23}=t$.
The peripheral QD1 and QD3 are symmetric and both in the localized momentum regime with $N_1=N_3=1$, respectively.
While QD2 is nearly unoccupied as $\varepsilon_2 > \varepsilon_F$.
The configuration of the TQDs device is set as $(N_{1},N_{2},N_{3})=(1,0,1)$, where $N_{i}$ is the number of electrons in $i$ dot.
The dot spectral density and the spatial spin correlation function will suggest a spectroscopic way of detecting the Kondo clouds.
In Ref.\cite{2017prb}, we have demonstrated a long-range antiferromagnetic exchange interaction in the TQDs system in the Kondo regime.
In the present work, as depicted in Fig.~\ref{fig0} (b),  we will present a clear physical picture of the long-range entanglement of Kondo clouds,
via study the transition of the spectral functions, spin-spin correlation and dot occupancies of the OTQDs.
This spatial extent of Kondo clouds in OTQDs system might be observed experimentally.

\section{MODEL AND DEOM THEORY}
In the geometry depicted in Fig.~\ref{fig0} (a), two peripheral quantum dots (QD-1 and QD-3) are two magnetic impurities of spin-$1/2$ directly coupled to the leads via hybridization widths $\Delta$, while the intermediate QD' energy level is high than Fermi level, as an empty orbital. $t$ is the interdot coupling strengths between the QD-1(3) and QD-2, determined by overlapping integral of them. The total Hamiltonian for the system is
\begin{align}\label{hs2}
H=H_{TQDs}+H_{leads}+H_{coupling}
  \end{align}
where the interacting TQDs part is described by three-impurity Anderson model
\begin{align}\label{hs2}
   H_{TQDs}=\sum_{\sigma i=1,2,3}[\epsilon_{i\sigma}\hat{a}^\dag_{i\sigma}\hat{a}_{i\sigma} + U_{i} n_{i\sigma}n_{i\bar{\sigma}}]
 \nl    + t \sum_{\sigma}( \hat{a}^\dag_{1\sigma}\hat{a}_{2\sigma}+\hat{a}^\dag_{2\sigma}\hat{a}_{3\sigma}+\text{H.c.})
  \end{align}
here $\hat{a}_{i\sigma}^\dag$ ($\hat{a}_{i\sigma}$) is the operator that creates (annihilates) a spin-$\sigma$ electron with energy $\epsilon_{i\sigma}$ ($i=1,2,3$) in $i$ th dot. $U_{i}$ is the on-dot Coulomb interaction between electrons with spin $\sigma$ and $\bar{\sigma}$ (opposite spin of $\sigma$). $n_{i\sigma}=\hat{a}^\dag_{i\sigma}\hat{a}_{i\sigma}$ is the operator of the electron number for the dot $i$.

The device leads can be treated as single-particle reservoirs and the Hamiltonian is
$H_{leads}=\sum_{k\mu\alpha=L,R}\epsilon_{k\alpha}\hat{d}^\dag_{k\mu\alpha}\hat{d}_{k\mu\alpha}$,
here we adopt the symbol $\mu$ to denote the electron orbital (including spin, space, \emph{etc.}) in the above Hamiltonian for brevity, i.e.,  $\mu=\{{\sigma},i...\}$.
$\epsilon_{k\alpha}$ is the energy of an electron with wave vector $k$ in the $\alpha$ lead, and $\hat{d}^\dag_{k\mu\alpha}$($\hat{d}_{k\mu\alpha}$) corresponds creation (annihilation) operator for an electron with the $\alpha$-reservoir state $|k\rangle$ of energy $\epsilon_{k\alpha}$.
The dot-lead coupling assumes the form of $H_{coupling}=\sum_{\mu\alpha}(\hat{a}^\dag_{\mu}\hat{F}^{-}_{\mu\alpha}+\hat{F}^{+}_{\mu\alpha}\hat{a}_{\mu})$, with $\hat{F}^{-}_{\mu\alpha}=\sum_{k}t_{k\mu\alpha}\hat{d}_{k\alpha}=(\hat{F}^{+}_{\mu\alpha})^{\dag}$ \cite{2014jcp054105}.
The influence of device leads on the TQDs are taken into account through the hybridization bath spectral density functions $J_{\alpha\mu\mu'}(\omega)=\pi\sum_{k} t_{k\mu\alpha}t^\ast_{k\mu'\alpha
} \delta(\w-\ep_{k\alpha})$, which can be defined via anticommutators as $J^{\varrho}_{\alpha\mu\mu'} \equiv 1/2[\int^{\infty}_{-\infty} dt\, e^{i\omega t}\langle\{\hat{F}^{\varrho}_{\mu\alpha}(t),\hat{F}^{\bar{\varrho}}_{\mu'\alpha}(0)\}\rangle_{B}]$.
Here, we have set $J_{\alpha\mu\mu'}(\w)\equiv J^{-}_{\alpha\mu\mu'}(\w) \equiv J^{+}_{\alpha\mu'\mu}(\w)$ and denote $\tilde{F}^{\varrho}_{\mu\alpha} \equiv \bar{\varrho} \hat{F}^{\varrho}_{\mu\alpha}$  with $\varrho=+,-$ and $\bar{\varrho}=-\varrho$ for identifying the creation/annihilation operators \cite{2015jcp234108}.

The interacting TQDs is accurately solved by the DEOM approach \cite{2014jcp054105}.
Established on the Feynman-Vernon path-integral formalism, DEOM is naturally a nonperturbative and accurate many-particle theory.
Based on the quasi-particle (dissipaton) description of hybridizing bath, the system-environment correlations are fully taken into consideration \cite{Jin08234703,2012prl266403}.
In principle, DEOM method can be used in the accurate evaluation on equilibrium and nonequilibrium properties of both system and hybridizing bath \cite{2015njp033009}.
The reduced dissipaton density operators of the quantum dot system $\rho^{(n)}_{j_1\cdots j_n}(t)\equiv {\rm tr}_{B}\,[(\hat{f}_{jn}\cdots \hat{f}_{j1})^{o}\rho_{\mathrm{T}}(t)]$ are the basic variables in DEOM.
Here $(\hat{f}_{jn}\cdots \hat{f}_{j1})^{o}$ specifies an ordered set of $n$ irreducible dissipatons and the so-called dissipaton operators $\hat{f}_{j}$ decomposes $\tilde{F}^{\varrho}_{\mu\alpha}$ as $\tilde{F}^{\varrho}_{\mu\alpha} \equiv \sum _{j}\hat{f}_{j}$ due to the exponential decomposition form of bath correlation function. We apply the Liouville-von Neumann equation,
\begin{align}\label{HEOM}
   {\dot{\rho}_{\mathrm{T}}(t)=-i[H_{dots}+H_{leads}+H_{coupling},\rho_{\mathrm{T}}(t)]}
\end{align}
for the total composite density operator $\rho^{(n)}_{j_1\cdots j_n}$. The DEOM formalism that governs the time evolution of dissipaton density operators can be obtained as \cite{2014jcp054105,2015jcp234108}:
\begin{align}\label{HEOM}
   \dot\rho^{(n)}_{j_1\cdots j_n} =& -\Big(i{\cal L} + \sum_{r=1}^n \gamma_{j_r}\Big)\rho^{(n)}_{j_1\cdots j_n}
%\nl &
     -i \sum_{j}\!     %\sideset{}{'}
     {\cal A}_{\bar j}\, \rho^{(n+1)}_{j_1\cdots j_nj}
\nl &
    -i \sum_{r=1}^{n}(-)^{n-r}\, {\cal C}_{j_r}\,
     \rho^{(n-1)}_{j_1\cdots j_{r-1}j_{r+1}\cdots j_n}
\end{align}
where, ${\cal A}_{\bar j}\equiv {\cal A}^{\bar{{\varrho}}}_{\alpha\mu m} = {\cal A}^{\bar{{\varrho}}}_{\mu}$ and ${\cal C}_{j_r} \equiv {\cal C}^{\varrho}_{\alpha\mu m}$ are Grassmannian superoperators, defined via
\begin{align}\label{HEOM}
   {\cal A}^{\varrho}_{\mu}\, \hat{O}_{\pm} \equiv \hat{a}^{\varrho}_{\mu} \hat{O}_{\pm}
\pm \hat{O}_{\pm} a^{\varrho}_{\mu} \equiv [\hat{a}^{\varrho}_{\mu}, \hat{O}_{\pm}]_{\pm}
\end{align}
\begin{align}\label{HEOM}
  {\cal C}^{\varrho}_{\alpha\mu m}\,\hat{O}_{\pm} \equiv \sum_{\mu'}(\eta^{\varrho}_{\alpha \mu \mu' m} \hat{a}^{\varrho}_{\mu'}\hat{O}_{\pm}
\mp \eta^{\bar{\varrho}^{*}}_{\alpha \mu \mu' m} \hat{O}_{\pm} \hat{a}^{\varrho}_{\mu'} )
\end{align}
Here, $\hat{O}_{\pm}$ denotes an arbitrary operator, with even (+) or odd (-) fermionic parity respectively and $\eta^{\varrho}_{\alpha \mu \mu' m}$ is the exponential bath correlation function components.
All $\{\rho^{(n)}_{j_1\cdots j_n}\}$ are now the physically well-defined dissipaton density operators. The hierarchical coupling can be simply truncated by setting all $\rho^{(n>L)}_{j_1\cdots j_n}=0$, at a sufficiently large $L$. While all $L$-body dissipatons dynamics are treated exactly, the resulting closed DEOM for $\{{\rho^{(n)}_{j_1\cdots j_n}};\,n=0,1,...,L\}$ represents also a dynamical meanfield scheme for higher-order dissipaton density operators.
Actually, it is analytically proved that DEOM formalism is exact at the hierarchy level $L= 8$ for single QDs, due to the underlying Grassman numbers properties.
As a non-perturbative theory, DEOM numerically converges rapidly with the Pad\'e spectrum decomposition (PSD) scheme.\cite{2012prl266403,2014jcp054105,2015jcp234108}
For the system studied in this work, DEOM evaluation is thought to be quantitatively accurate at $L= 4$.

In this sense, the physical quantities can be calculated via the DEOM-space linear response theory.
The QD occupancy is
\begin{align}\label{hd}
    n_{\mu}=\sum_{\mu} n_{\mu}=tr_{s}[\hat{a}^\dag_{\mu}\hat{a}_{\mu}\rho^\dag_{\alpha \mu}]
  \end{align}
The electric current operator from bath $\alpha$-lead to system is given by
\begin{align}\label{hd}
    \hat{I}_{\alpha}=-\frac{d}{dt}(\sum_{k}\hat{d}^\dag_{k\alpha}\hat{d}_{k\alpha})=
    -i\sum_{\mu}(\hat{a}^\dag_{\mu}\hat{F}^{-}_{\mu\alpha}-\hat{F}^{+}_{\mu\alpha}\hat{a}_{\mu})
  \end{align}%
The mean current can then be evaluated in terms of dissipaton density operators as $I_{\alpha}(t)=\mathrm{Tr_{T}}[\hat{I}_{\alpha} \rho_{\mathrm{T}}(t)]$.
The spectral function
\begin{align}\label{hd}
    A_{\mu}(\omega)=\frac{1}{\pi}Re\Big\{\int^{\infty}_{0}dt \{\tilde{\cal C}_{\hat{a}^\dag_{\mu}\hat{a}_{\mu}}(t)+
    [\tilde{\cal C}_{\hat{a}_{\mu}\hat{a}^\dag_{\mu}}(t)]^{\ast}\}e^{i\omega t}\Big\}
  \end{align}%
with the system correlation functions $\tilde{\cal C}_{\hat{a}^\dag_{\mu}\hat{a}_{\mu}}(t)$ and $\tilde{\cal C}_{\hat{a}_{\mu}\hat{a}^\dag_{\mu}}(t)$ following immediately the time-reversal symmetry and detailed-balance relations which are evaluated through the time evolution of the DEOM propagator. The details of the DEOM formalism and the derivation of physical quantities are supplied in the Refs.\cite{2014jcp054105,2015jcp234108,2012prl266403}.

\section{RESULTS AND DISCUSSION}

\begin{figure}
\includegraphics[width=0.95\columnwidth]{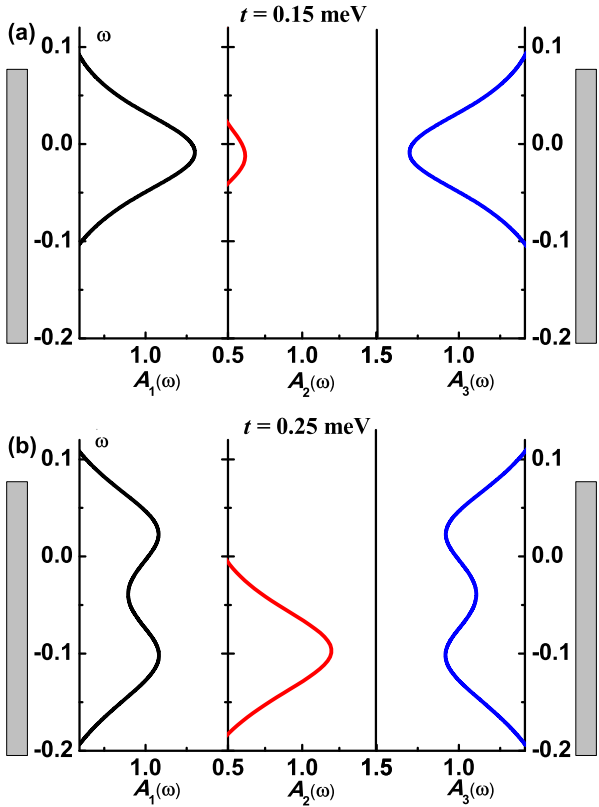}
\caption{(Color online) The spectral functions $A_i(\omega)$ of the OTQDs for the interdot coupling strength $t = 0.15\mathrm{meV}$ (a) and $t = 0.25\mathrm{meV}$ (b).
The parameters adopted are $\epsilon_{\mathrm{1,3}} = -0.6\mathrm{meV}$, $U_{\mathrm{1,3}} = 1.2\mathrm{meV}$, $\epsilon_{\mathrm{2}} = 0.3\mathrm{meV}$, $U_{\mathrm{2}} = 0\mathrm{meV}$, $W = 5.0\mathrm{meV}$, $\Delta = 0.3\mathrm{meV}$, $K_{B}T = 0.03\mathrm{meV}$. The Kondo temperature for those parameters is $K_{B}T_K = 0.13\mathrm{meV}$.}
\label{fig1}
\end{figure}

We present the numerical solution of the OTQDs model in Fig. \ref{fig0} (a) using the DEOM method.
For simplicity, we assume that the QD1 and QD3 are same and thus have the same parameters: $\epsilon_{\mathrm{1,3}} = -0.6\mathrm{meV}$, $U_{\mathrm{1,3}} = 1.2\mathrm{meV}$, $\epsilon_{\mathrm{2}} = 0.3\mathrm{meV}$, $U_{\mathrm{2}} = 0\mathrm{meV}$.
The QD1 and QD3 always possesses electron-hole symmetry.
We set $W = 5.0\mathrm{meV}$ and $\Delta = 0.3\mathrm{meV}$ for lead bandwidth and  dot-lead coupling strength, respectively.
According to a theoretical estimation $T_K = \sqrt{\frac{U\Delta}{2}}e^{-\pi U/8\Delta +\pi \Delta/2U}$ \cite{1993cambrige}, we calculate the Kondo temperature is $K_{B}T_K = 0.13\mathrm{meV}$.
Here, we adopt a low temperature $K_{B}T = 0.03\mathrm{meV}$ to ensure that the OTQDs system remains in Kondo regime.
Firstly, we calculate the spectral functions $A_i(\omega)$ of the three QDs for the interdot coupling strength $t = 0.15\mathrm{meV}$.
The numerical result are depicted in Fig. \ref{fig1} (a).
We find that the spectral functions both for QD1 and QD3 exhibit a single Kondo peak for the interdot coupling strength $t = 0.15\mathrm{meV}$.
Here, just as the Kondo state of single QD, the QD1(3) prefers to form its respective Kondo singlet state with the conduction electrons of left (right) lead.
More interesting issue is that for the empty orbital QD2 ($\epsilon_{\mathrm{2}} = 0.3\mathrm{meV}$), the spectral function also arises a single peak.
This peak results from the overlapping of the two Kondo clouds forming from between the QD1 and left lead and the QD3 and the right lead.
It is a long-range entanglement of Kondo quasiparticles localized in QD1 and QD3, as sketched in Fig. \ref{fig0} (b).
To confirm the peak of QD2' spectral function is the conduction electrons peak.
We also increase the interdot coupling strength to $t = 0.25\mathrm{meV}$, and present the spectral functions $A_i(\omega)$ of the three QDs in Fig. \ref{fig1} (b).
Due to the long-range effective antiferromagnetic exchange interaction between QD1 and QD3, the single Kondo peak of QD1(3) will splits into two peaks.
Here, the electron peak and hole peak are separated from each other.
The electron peak will be localized at $\omega < 0$.
While the position of the hole peak is above the Fermi level.
It is found that the peak of spectral function for QD2 also moves to the below of Fermi level, corresponding to the position of the electron peak of QD1(3).
More importantly, the conduction electrons peak of QD2 becomes higher and broadens with the interdot coupling strength $t$.
The reason is that the strong interdot coupling will enhance the electron's hopping between the three QDs.

\begin{figure}
\includegraphics[width=0.95\columnwidth]{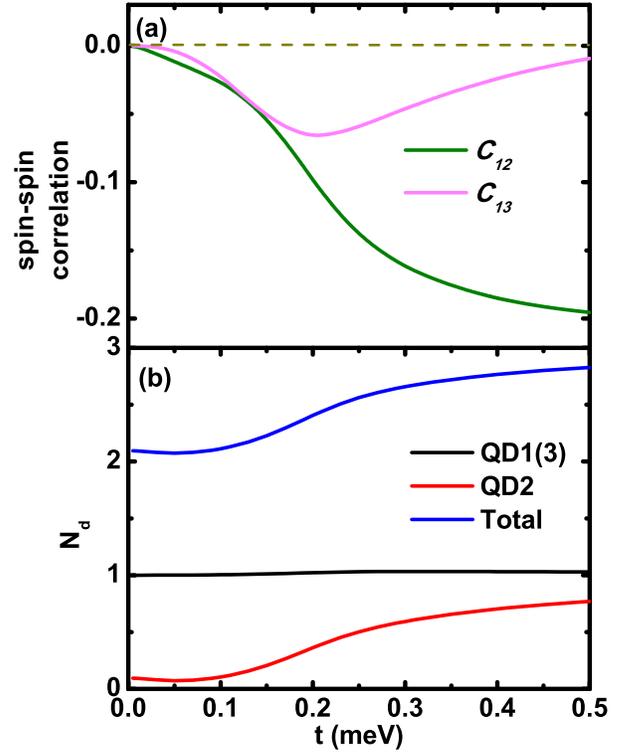}
\caption{(Color online) (a) The spin-spin correlation $C_{ij}$ of the OTQDs versus the interdot coupling strength $t$.
(b) The corresponding electron occupation $N_{d}$ of the three QDs and the total electron occupation of the OTQDs versus the interdot coupling strength $t$.
Due to the overlapping of the Kondo clouds forming from the QD1 and left lead and the QD3 and the right lead, an effective electron occupation exhibits on QD2.
However, the electron occupations of QD1 and QD3 keep in $N_{d} = 1$.
The parameters are the same as those in \Fig{fig1}.}
\label{fig2}
\end{figure}
The above physical picture for the long-range entanglement of Kondo clouds can also be illustrated by the transformation of the spin-spin correlation and electron occupation of OTQDs system.
we calculate the spin-spin correlation function between QD $i$ and $j$ as,
\begin{equation}\label{SSCF}
 C_{ij}\equiv\langle \vec{S}_{i}\cdot \vec{S}_{j} \rangle-\langle \vec{S}_{i}\rangle\cdot\langle \vec{S}_{j}\rangle.
\end{equation}
We investigate the transition of the spin-spin correlation between the two adjacent quantum dots (QD1 and QD2) $C_{12}$ and between the two peripheral quantum dots (QD1 and QD3)$C_{13}$ and the transition of the electron occupation $N_d$ for the three QDs in Kondo regime.
The numerical results are shown in Fig. \ref{fig2}.
It is noticed that both $C_{12}$ and  $C_{13}$ are negative.
This indicates that there are two antiferromagnetic interactions in the OTQDs.
One is the long-range effective antiferromagnetic exchange interaction between QD1 and QD3, associated with $C_{13} < 0$ \cite{2017prb}.
The another antiferromagnetic interaction between QD1 and QD2 results from the overlapping between the left Kondo clouds forming from the QD1 and left lead and the right Kondo clouds forming from the QD3 and right lead.
With increasing interdot coupling strength $t$, the spin-spin correlation between the two adjacent QD1 and QD2 $C_{12}$  becomes stronger and gradually achieves a finite value.
The larger $C_{12}$ indicates an enhancement of the entanglement of the left and right Kondo clouds.
This also can be illustrated by the electron occupation $N_d$.
Figure \ref{fig2} (b) presents the interdot coupling strength dependent electron occupation $N_d$ of the three QDs.
It is noticed that the electron occupation $N_d$ of QD2 increases firstly with the interdot coupling $t$ and finally reaches a steady state value.
The effective electron occupation exhibits on empty orbital QD2 also indicates the overlapping of the two Kondo clouds.
On the other hand, the spin-spin correlation between the two peripheral QDs (QD1 and QD3)$C_{13}$ shows an nonmonotonic transition behaviour.
It firstly increases at low interdot couplings ($t < 0.25\mathrm{meV}$) and then decreases with the stronger interdot coupling strength ($t > 0.25\mathrm{meV}$).
This is due to there is a transition from the degenerate Kondo singlet states of individual QDs at weak interdot coupling strength to long-range singlet state developing from QD1 and QD3 at strong interdot coupling strength  \cite {2017prb}.
The competition between the degenerate Kondo singlet states and long-range singlet state leading the nonmonotonic variation of long-range spin-spin correlation $C_{13}$.
However, the electron occupation of QD1(3) always keep in $N_d =1$.
As an open system, the conduction electrons of the leads implant in the TQDs due to the long-range entanglement of Kondo clouds£¬
,which leads to the total electron occupation $N_d$ of the TQDs increasing with the interdot coupling strength.

\begin{figure}
\includegraphics[width=0.95\columnwidth]{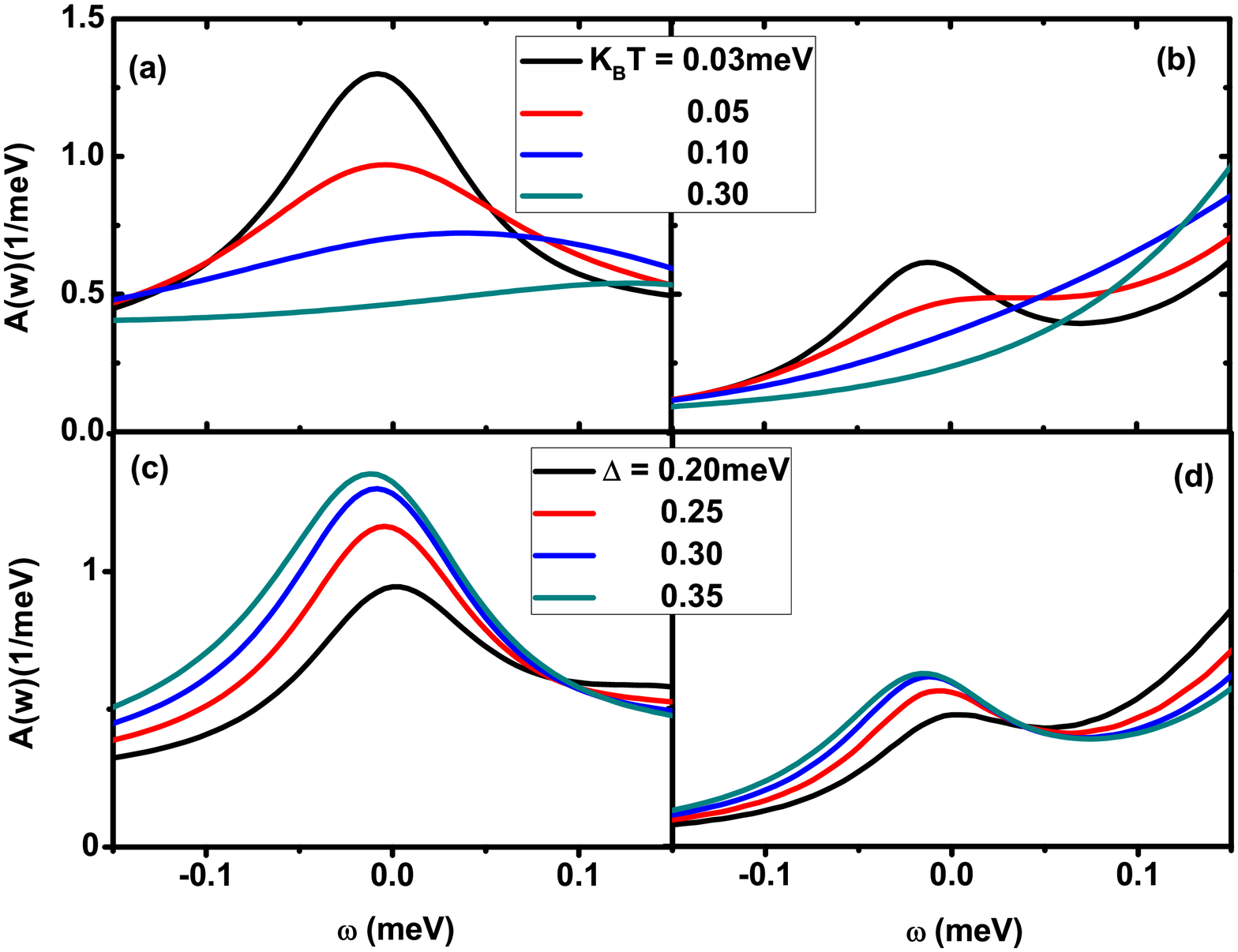}
\caption{(Color online) The spectral functions $A(\omega)$ of the QD1(3) (a) and QD2 (b) with different temperatures $K_{B}T$ for the interdot coupling strength $t = 0.15\mathrm{meV}$. The spectral functions $A(\omega)$ of the QD-1(3) (c) and QD-2 (d) with different dot-lead coupling strengths $\Delta$ for the interdot coupling strength $t = 0.15\mathrm{meV}$.
The other parameters are the same as those in \Fig{fig1}.}
\label{fig3}
\end{figure}
To further study thoroughly the physical phenomenon of the long-range entanglement of Kondo clouds, we also calculate the spectral functions $A(\omega)$, spin-spin correlation  $C_{12}$ and  $C_{13}$ and electron occupation $N_{di}$ with the influence of temperature and dot-lead coupling strength.
The temperature dependent spectral functions $A(\omega)$ of QD1(3) and QD2 for the interdot coupling strength $t = 0.15\mathrm{meV}$ are shown in Fig. \ref{fig3} (a) and (b), respectively.
We find that the Kondo single peak of QD1(3) is high and sharp at low temperatures.
With increasing temperature, the height of Kondo peak progressively decreases and finally vanishes at temperature  $T > T_K$ (see Fig. \ref{fig3} (a)).
The temperature plays an inhibitory effect on the Kondo physics of the OTQDs system.
The enhancement of Kondo effect at low temperature will strengthen the conduction electrons peak of spectral function for QD2.
So the peak of spectral function for QD2 also increases with decreasing the temperature.
When the temperature is higher than the Kondo temperature $T_K$, the suppression of Kondo effect will leads to that there is no conduction electrons peak on QD2 (see the dark cyan curve at $K_{B}T = 0.3\mathrm{meV}$ in Fig. \ref{fig3} (b)).
More over, as an effect of the overlapping of Kondo clouds, the height of the conduction electrons peak of QD2 is much lower than the Kondo single peak of QD1 at the same temperature.

We then examine the dot-lead coupling strength dependent long-range entanglement of Kondo clouds.
Figure \ref{fig3} (c) and (d) show the spectral functions $A(\omega)$  of the QD1(3) and QD2 for the interdot coupling strength $t = 0.15\mathrm{meV}$.
We find that the height of the single Kondo peak of QD1(3) rises with increasing dot-lead coupling strength $\Delta$.
According to the analytical expression for the Kondo temperature of the single QD system $T_K = \sqrt{\frac{U\Delta}{2}}e^{-\pi U/8\Delta +\pi \Delta/2U}$ \cite{1993cambrige},
the Kondo temperature $T_K$ increases with the dot-lead coupling strength $\Delta$.
Since the temperature of the OTQDs system is fixed at $K_BT = 0.03\mathrm{meV}$,
this leads to the Kondo effect enhancing with the increasing of $\Delta$, accompanied with the rising height of the Kondo peak (see Fig. \ref{fig3} (c)).
Importantly, the enhancement of Kondo effect will expand the space of Kondo clouds, which leads to that
the conduction electrons peak of QD2 resulting from the overlapping of two Kondo clouds also becomes higher with the strong dot-lead couplings (see Fig. \ref{fig3} (d)).
\begin{figure}
\includegraphics[width=0.95\columnwidth]{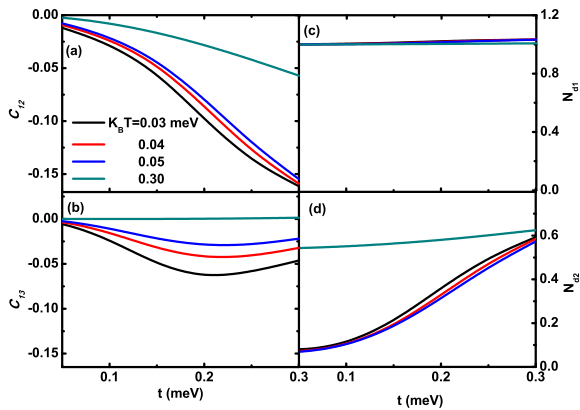}
\caption{(Color online) The spin-spin correlation $C_{12}$ (a) and $C_{13}$ (b) as a function of the interdot coupling strength $t$ with the different temperature. (c) and (d) is the corresponding electron occupancy $N_{d1}$ for QD1(3) and $N_{d2}$ for QD2, respectively.
The other parameters are the same as those in \Fig{fig1}.}
\label{fig4}
\end{figure}

As the long-range quantum entanglement of Kondo clouds, the temperature plays an very important part on this phenomenon.
We proceed with studying the impact of the temperature on the transition of the spin-spin correlation and dot occupancy.
The temperature dependent spin-spin correlation $C_{12}$ and $C_{13}$ as a function of the interdot coupling strength $t$ are shown in Fig. \ref{fig4} (a) and Fig. \ref{fig4} (b), respectively.
And the corresponding dot occupancies $N_{d1}$ and $N_{d2}$ are presented in Fig. \ref{fig4} (c) and Fig. \ref{fig4} (d), respectively.
We find that at low temperature (such as $K_{B}T = 0.03\mathrm{meV}$), both the spin-spin correlations $C_{12}$ and $C_{13}$ is distinct.
With the temperature increasing, the two spin-spin correlations are become attenuated.
At very high temperature (such as $K_{B}T = 0.3\mathrm{meV}$), the spin-spin correlation $C_{13}$ becomes vanished.
Only, the finite spin-spin correlation $C_{12}$ is conserved in the OTQDs system.
The reason is that at low temperature, the enhancement of the Kondo effect will expands the space of the Kondo clouds,which leads to
the effective dot occupancy of conduction electrons on the intermediate QD2 is large (see Fig. \ref{fig4} (d)).
So the spin-spin correlation $C_{12}$ increase with the decreasing temperature.
However, the dot occupancy of QD1(3) almost maintains $N_{d1} = 1$ for all the temperatures (see Fig. \ref{fig4} (c)).
The long-range effective antiferromagnetic exchange interaction between QD1 and QD3 correlated Kondo effect causes the long-range spin-spin correlation $C_{13}$ firstly increasing and then achieving a finite value with the interdot coupling $t$.
Moreover, for the same finite intedot coupling strength, the spin-spin correlations $C_{12} > C_{13}$ indicates the antiferromagnetic spin coupling between QD1 and QD2 stronger than the long-range effective antiferromagnetic exchange interaction between QD1 and QD3.
At very high temperature (such as $K_{B}T = 0.3\mathrm{meV}$), the OTQDs system is out of the Kondo regime, which leads to
that the long-range effective antiferromagnetic exchange interaction between QD1 and QD3 correlated Kondo effect becomes vanished as $C_{13} = 0$.
The charge fluctuation associated with high temperature plays a more major part in OTQDs system.
It results in a very lager dot occupancy of QD2 than Kondo regime.
So an effective spin-spin correlation $C_{12}$ also maintains in the OTQDs system.
\begin{figure}
\includegraphics[width=0.95\columnwidth]{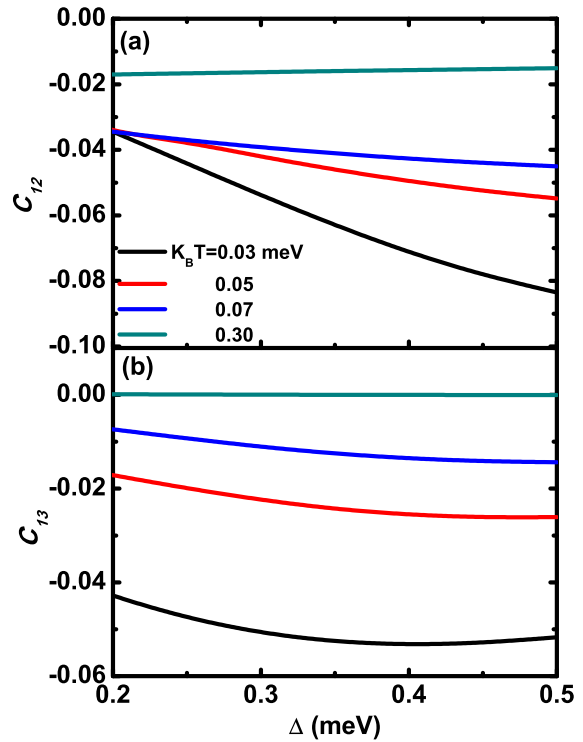}
\caption{(Color online)  The spin-spin correlation $C_{12}$ (a) and $C_{13}$ (b) as a function of the dot-lead coupling strength $\Delta$ with the different temperature for the interdot coupling $t=0.15$ meV.
The other parameters are the same as those in \Fig{fig1}.}
\label{fig5}
\end{figure}

Finally, we elucidate the spin-spin correlation $C_{12}$ and $C_{13}$ as a function of the dot-lead coupling strength $\Delta$ with different temperature.
The results are presented in Fig. \ref{fig5} (a) and (b), respectively.
It is noticed that the spin-spin correlation $C_{12}$ increases monotonously with the dot-lead coupling strength $\Delta$ in Kondo regime.
The strong dot-lead coupling strength $\Delta$ will causes an enhancement of screening on the spin of QD1(3) by the conduction electrons of leads.
So the long-range entanglement of Kondo clouds also enhances with the strong dot-lead coupling strength $\Delta$.
With the temperature increasing, the Kondo effect of OTQDs system is weakened.
The spin-spin correlation $C_{12}$ associated with the overlapping of Kondo clouds is subdued by the high temperature (see Fig. \ref{fig5} (a)).
The dot-lead coupling strength $\Delta$  plays a weaker effect on the long-range spin-spin correlation $C_{13}$.
With increasing the dot-lead coupling strength $\Delta$, the long-range spin-spin correlation $C_{13}$ firstly increases with weak dot-lead coupling strength $\Delta$ and then decreases with strong dot-lead coupling strength $\Delta$.
This is due to the fact that there is a transition from the degenerate Kondo singlet states to long-range singlet state in the OTQDs system.
The competition between the degenerate Kondo singlet state and long-range singlet state leading the nonmonotonic variation of long-range spin-spin correlation $C_{13}$.
In general, both the spin-spin correlation $C_{12}$ and $C_{13}$ related to the Kondo physics are suppressed by the temperature.

\section{CONCLUSIONS}

In summary, a clear picture of the long-range entanglement of Kondo clouds in OTQDs system is described.
We present an overlapping of the Kondo clouds phenomenon forming from the QD1 and left lead and the QD3 and right lead by investigating the spectral functions, spin-spin correlation and dot occupancies.
A conduction electrons peak of the spectral function is found on the empty orbital QD2 in Kondo regime, which resulting from the overlapping of the Kondo clouds enhances with the low temperature and strong dot-lead coupling strength.
The negative spin-spin correlation $C_{12}$ and $C_{13}$ of OTQDs and an effective electron occupation on QD2 exhibiting the classical temperature $T$ dependence also indicates this long-range entanglement phenomenon of Kondo quasiparticles in OTQDs.
It will be valuable to find and confirm experimentally the entanglement of Kondo clouds.
Most important is that, we present a direct access to the manipulation on long-range entanglement of Kondo clouds in the many-body quantum system, which is very significant for quantum transport and communication.

\section{ACKNOWLEDGMENT}

This work was supported by the NSF of China (Grants No.\,11804245, No.\,11747098,  No.\,11774418, No. \,11674139, No. \,11834005, No. \,U1530401) and Program for the Innovative Talents of Taiyuan Institute of Technology.
%

%\bibliographystyle{aip}
%\bibliography{bibrefs}

%\iffalse

%\fi

\end{document}